\documentclass[a4paper,11pt]{article}
\usepackage{pos}
\usepackage{placeins} 
\usepackage{graphicx}
\graphicspath{ {./figures/} }

\title{The negligible impact of experimental inconsistencies in the NNPDF4.0 global dataset}

\author*{Roy Stegeman}

\affiliation{TIF Lab, Dipartimento di Fisica, Universit\`a degli Studi di
  Milano and INFN Sezione di Milano \\
  Via Celoria 16, 20133, Milano, Italy
}

\emailAdd{r.stegeman@mi.infn.it}

\abstract{As both predictions and measurements of high-energy physics observables become more precise, controlling all sources of uncertainties in determinations of parton distribution functions (PDFs) becomes increasingly important. One source of PDF uncertainty is the result of data not being consistent under a chosen theoretical framework. In these proceedings we investigate the impact these inconsistencies present in the global NNPDF4.0 dataset. We show that, when accounting for missing higher order uncertainties, the missing contribution to the PDF uncertainty due to data inconsistencies are at the level of statistical fluctuations.
}

\FullConference{%
  41st International Conference on High Energy physics - ICHEP2022\\
  6-13 July, 2022\\
  Bologna, Italy
}

\begin{document}
\maketitle

\section{Inconsistencies between experimental measurements}
%

In the NNPDF4.0~\cite{Ball:2021leu} determination of parton distribution functions (PDFs) the obtained chi-squared statistic, $\chi^2$, representing the agreement to the experimental data is 1.16 per datapoint. With over 4000 datapoints in the global dataset it is exceedingly unlikely that this deviation from 1 is the result of a statistical fluctuation, instead the likely explanation is that this is because the data is not fully consistent when combined in the theoretical framework of perturbative QCD at next-to-next-to leading order (NNLO) used to extract the PDFs from data. Part of this observed inconsistency is certainly due to the fact that in the NNPDF4.0 determination missing higher order uncertainties (MHOUs) are not accounted for. MHOUs can however be included through a scale variations procedure~\cite{NNPDF:2019vjt}, and in these proceedings we will estimate their impact on the observed inconsistencies. Then, there may remain further experimental inconsistencies after accounting for MHOUs which are beyond the control of a PDF fitter.
%
%
To account for various sources of uncertainty, including that related to the impact of possible data inconsistencies, some PDF fitting collaborations therefore rely on a ``tolerance'' method in which the $\chi^2$ variations are inflated by a factor of around 5. In these proceedings we will show that, if MHOUs are taken into account, the residual uncertainty due to inconsistencies are at the level of statistical fluctuations, suggesting that at the current level of PDF precision no further procedure is required to account for them.

\section{Weighted fits and closure test data}
To test whether a given subset of data is affected by inconsistencies, one can perform a PDF fit while enforcing a high level of agreement with this specific subset of data by giving it a larger weight in the fit. If there are no inconsistencies present, this should be possible without causing the agreement with any of data present in the NNPDF4.0 global dataset to deteriorate, while improving the agreement to the dataset that is given additional weight. However, if inconsistencies (either within the subset of data or with respect to other data) are present, enforcing a high level of agreement will emphasize their effect in the form of increased PDF uncertainty. More details on the various scenarios that may occur in fits with increased weight in the presence of data inconsistencies can be found in Sect.~4.2 of Ref.~\cite{Ball:2021leu}.

Weighted fits, in which subsets of data are given a larger weight than justified by the experimental covariance matrix, were first used in Ref.~\cite{Forte:2020pyp} to study the determination of the strong coupling $\alpha_s(M_Z)$, and later in Ref.~\cite{Ball:2021leu} to check for inconsistencies in candidate datasets considered for the NNPDF4.0 global PDF determination. Below we will explain why this check, however, cannot guarantee all inconsistent datasets to be identified.
In a weighted fit the $\chi^2$ (defined using the $t_0$ prescription~\cite{Ball:2009qv} to avoid D'Agostini bias~\cite{D'Agostini:642515}) is adjusted to increase the effective number of datapoints in the subset to be half the total number of datapoints:
\begin{equation}
  \chi^2=\frac{1}{N_{\text {dat }}} \sum_{i=1}^{N_{\rm proc }} N_{\text {dat }}^{(i)} \chi_i^2 \quad \longrightarrow \quad \chi^2=\frac{1}{N_{\text {dat }}-N_{\text {dat }}^{(j)}} \sum_{i \neq j}^{N_{\rm proc }} N_{\text {dat }}^{(i)} \chi_i^2+\omega^{(j)} \chi_j^2,
  \label{eq:redefine_chi2}
\end{equation}
where $N_{\rm dat}^{(i)}$ is the number of datapoints in the subset of data $i$, and $\chi^2_i$ the figure of merit calculated to the corresponding subset of datapoints. The weights, $\omega^{(j)}$, are given by
\footnote{Note that the definition used in Ref.~\cite{Ball:2021leu} is $\omega^{(j)}={N_{\text {dat }}}/{N_{\text {dat }}^{(j)}}$, which does not correspond to exactly half the effective number of datapoints. For this study the difference is not relevant since inconsistencies would show using either definition as explicitly checked in Ref.~\cite{Ball:2021leu} by repeating the exercise while doubling and halving the values of $\omega^{(j)}$.}
\begin{equation}
  \omega^{(j)}=\frac{N_{\text {dat }}}{N_{\text {dat }}^{(j)}}-1.
\end{equation}

To understand why enforcing good agreement with a subset of data leads to an increased PDF uncertainty, consider that individual datapoints are subject to statistical fluctuations which will be overestimated if a datapoint is given a large weight. As an extreme example one may consider the case of a single datapoint, if this is given a large weight it will always be fitted with good agreement. However, this is clearly true regardless of whether data inconsistencies are present. To separate the effect due to this statistical fluctuation from the effect due to data inconsistencies, we compare the impact of weighted fits in two settings. First, the experimental setting in which we perform a fit to the experimentally measured values, in this setting our results are susceptible to any inconsistencies between measurements. Second, the closure test setting in which we perform fits to closure test data which is perfectly consistent by construction, thus allowing us to assess the increased uncertainty in the weighted fit procedure that is purely due to these statistical fluctuations. The check for inconsistent datasets performed in Ref.~\cite{Ball:2021leu} did not explicitly evaluate the effect due to these statistical fluctuations, therefore it might be that the NNPDF4.0 global dataset contains datasets for which the impact of a weighted fit is larger than justified based on statistical fluctuations (though the impact would still be limited, since those datasets corresponding to a large deterioration have indeed been discarded).

A closure test dataset is free of inconsistencies because it is generated from the theoretical predictions of a known PDF while the artificial statistical fluctuations are accounted for by sampling the experimental covariance matrices. The PDF resulting from a fit to closure test data thus reproduces the PDF used to generate the closure test data, rather than approximating the true proton PDF. It should be noted that data generated by sampling the covariance matrix in this way is generally referred to as ``level-2 closure test data'' in NNPDF publications.
%
%
For more details about closure tests the reader is referred to Refs.~\cite{Ball:2014uwa,DelDebbio:2021whr}.
%

The main objective of these proceedings is to compare the increase in uncertainty between the closure test setting and the experimental setting, this will allow us to gain insight into the magnitude of the missing contribution to the PDF uncertainties due to inconsistencies in the experimental data.
To quantify the PDF uncertainty, we use the $\varphi$ estimator introduced in Ref.~\cite{Ball:2014uwa}:
\begin{equation}
  \varphi \equiv \sqrt{\left\langle\chi^2\left[ T^{(k)}\right]\right\rangle-\chi^2\left[\left\langle T^{(k)}\right\rangle\right]}
  \label{eq:phi}
\end{equation}
where $T^{(k)}$ is the theoretical prediction corresponding to the $k$-th PDF replica, $f^{(k)}$, and the angled brackets denote the usual average over replicas for an observable $\mathcal{O}$ as
\begin{equation}
  \left\langle \mathcal{O}\left[f\right]\right\rangle=\frac{1}{N_{\text {rep }}} \sum_{k=1}^{N_{\text {rep }}} \mathcal{O}\left[f^{(k)}\right].
\end{equation}
The $\varphi$ indicator measures the standard deviation over the replica sample in units of the data uncertainty as is explicitly shown in Ref.~\cite{Ball:2014uwa}.

\section{Data inconsistencies in the NNPDF4.0 global dataset}
\begin{figure}
  \centering
  \includegraphics[width=.6\textwidth]{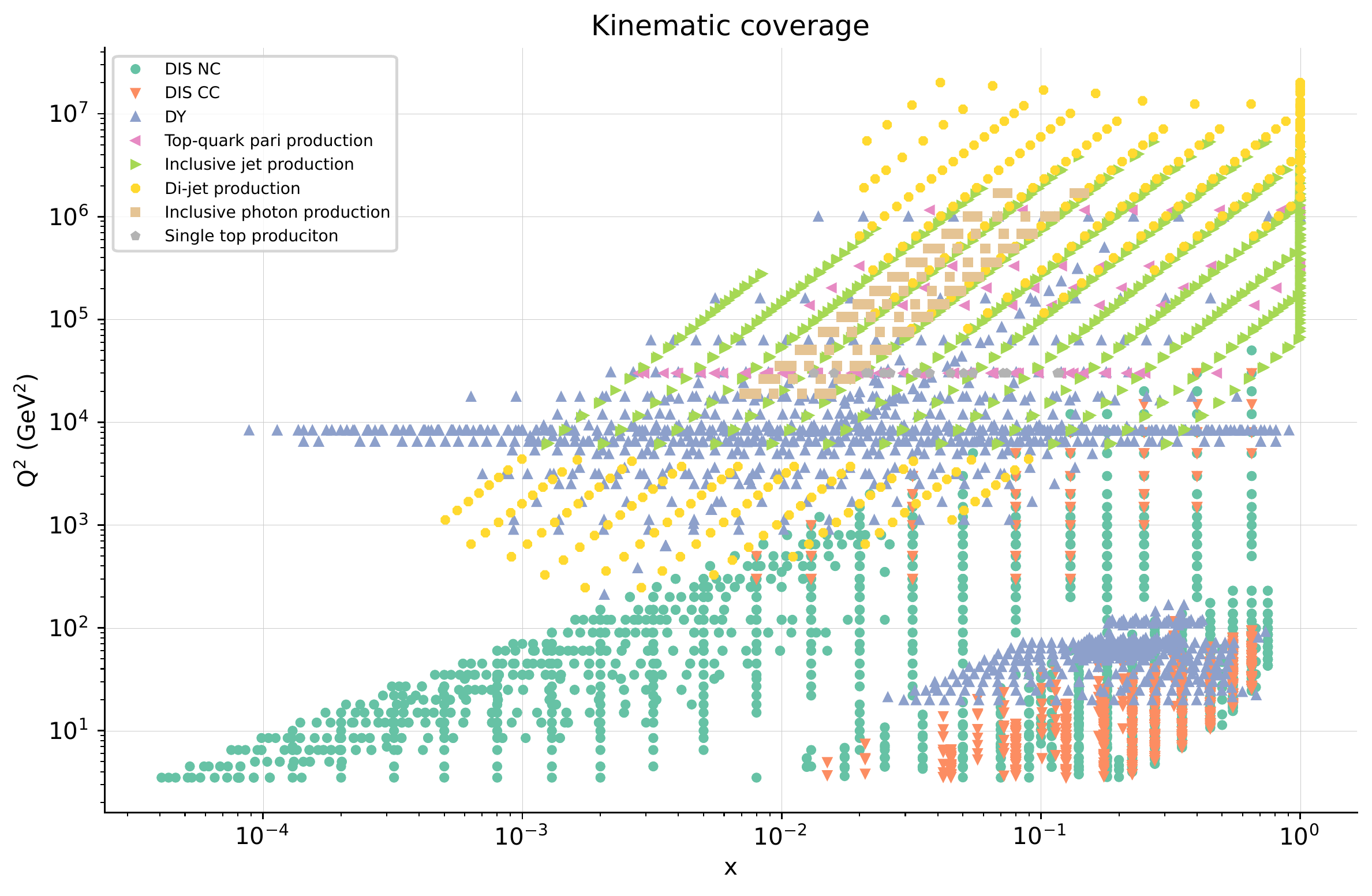}
  \caption{The kinematic coverage in the $(x,Q^2)$ plane of the NNPDF4.0 global dataset. Datasets have been labeled by process. This figure and others in these proceedings have been generated using \texttt{Reportengine}~\cite{zahari_kassabov_2019_2571601}.}
  \label{fig:kinplot}
\end{figure}

The weighted fits are produced by dividing the NNPDF4.0 global dataset as shown in appendix~B of Ref.~\cite{Ball:2021leu} into subsets based on the corresponding process and characterizing them as one of the following: Drell-Yan, (DY) neutral current (NC) deep inelastic scattering (DIS), charged current (CC) DIS, direct photon production, top production, single-inclusive jet production, di-jet production, and single top production. The kinematic coverage in the $(x,Q^2)$ plane of the datasets included in the NNPDF4.0 determination, subdivided into groups corresponding to the previously mentioned processes, are shown in Fig~\ref{fig:kinplot}. For hadronic observables the corresponding kinematic variables have been determined based on leading order (LO) kinematics, and for observables integrated over rapidity the corresponding $x$ value is determined using the center of the integration domain.

\begin{figure}
  \centering
  \includegraphics[width=.49\textwidth]{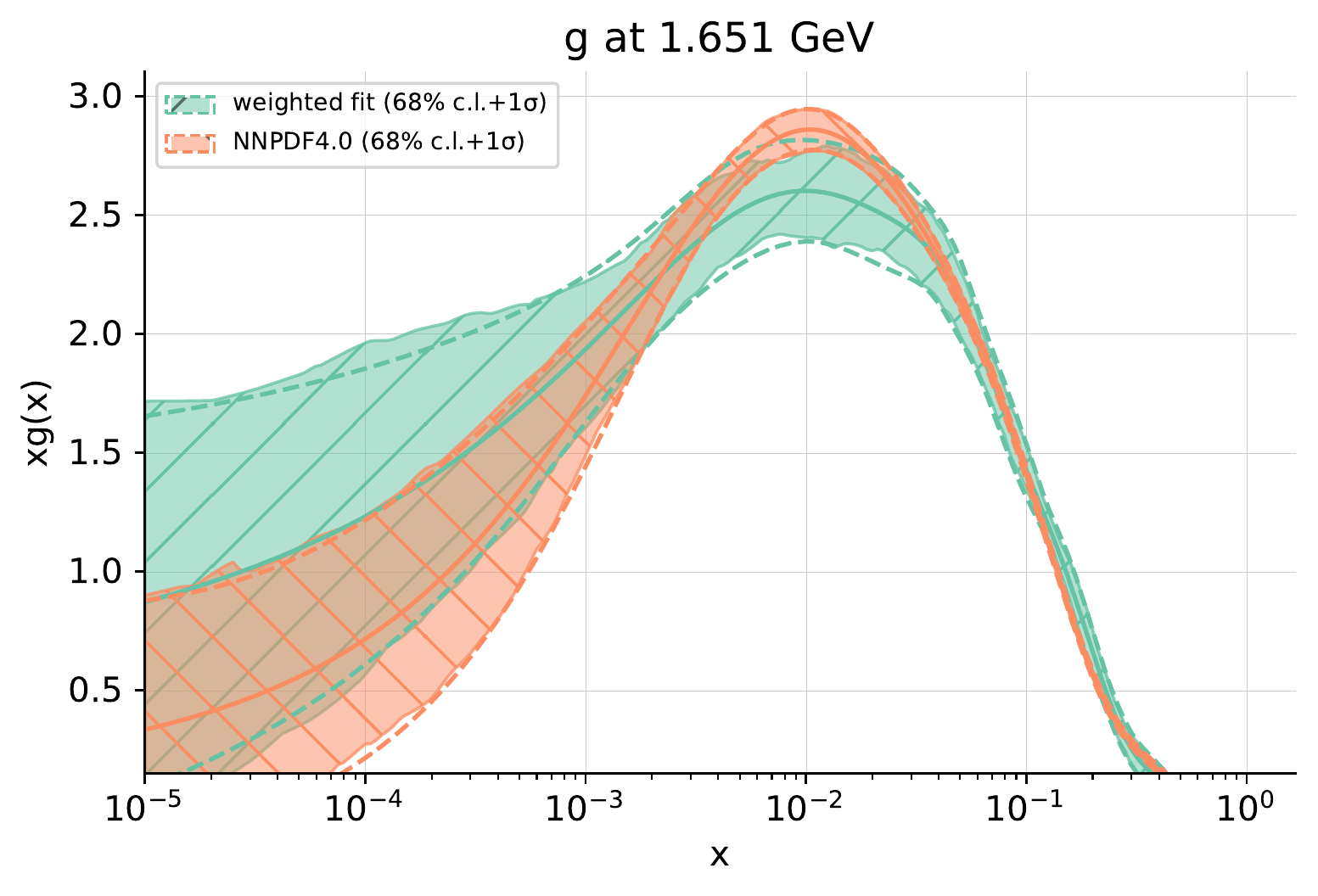}
  \includegraphics[width=.49\textwidth]{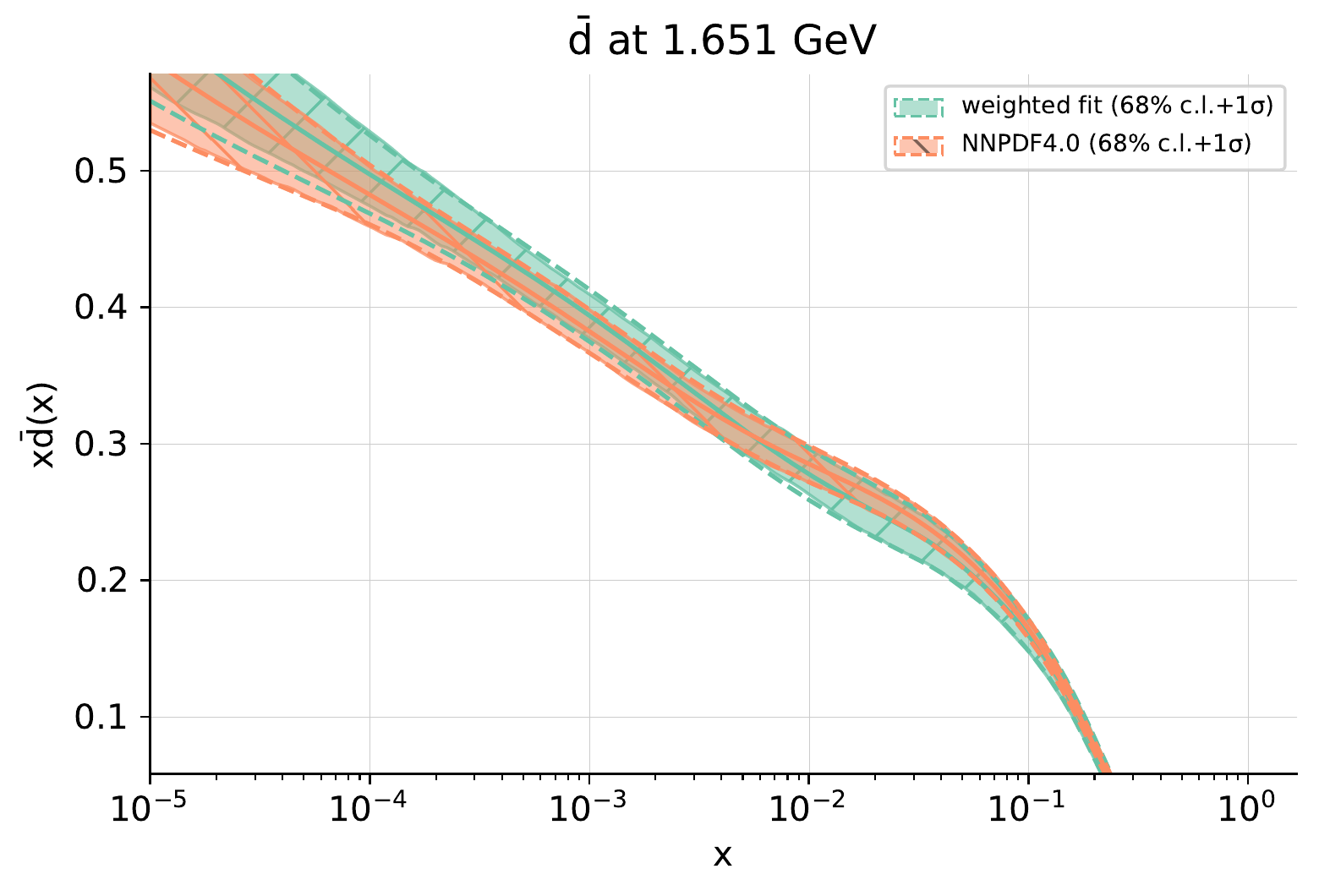}
  \caption{A comparison between the NNPDF4.0 baseline fit (orange) and a fit with datasets corresponding to top production weighted according to Eq.~(\ref{eq:redefine_chi2}) (green) for both the gluon PDF (left) and the anti-down PDF (right) at $Q=1.651$~GeV.}
  \label{fig:weighted_fit_pdf_plot}
\end{figure}

Upon giving additional weight to a subset of the data per Eq.~(\ref{eq:redefine_chi2}), in particular the gluon uncertainties are affected. To illustrate this, Fig.~\ref{fig:weighted_fit_pdf_plot} shows a comparison between the gluon and anti-down PDFs of the NNPDF4.0 baseline determination and a fit in which the datasets corresponding to measurements of top production have been given additional weight according to Eq.~(\ref{eq:redefine_chi2}). It is clear that the central value of the gluon is shifted by a large amount, though it is important to note that this shift is compatible with the increased uncertainties. The central value of the anti-down PDF, on the other hand, has only shifted by a small amount, and correspondingly the change in uncertainty is smaller.

In Tab.~\ref{tab:phivalues} the PDF uncertainties corresponding to the weighted fits for each process expressed in terms of $\varphi$, Eq.~(\ref{eq:phi}), are shown. The first column lists the relevant process, the second column the total number of datapoints present in the corresponding datasets, and the third column shows the results for weighted fits performed to experimental data (as opposed to closure test data).
%
%
Thus, for example, the uncertainty expressed in terms of the $\varphi$ value for the NNPDF4.0 without any additional weight given to any subset of data is 0.1569±0.0041.
The rows below this value then show the $\varphi$ value of the different variations of weighted fits that have been performed. It can be seen that for processes with a smaller number of corresponding datapoints the $\phi$ value is larger.

\begin{table}[]
  \begin{tabular}{l|l|l|l|l}
    Process    & $N_{\rm dat}^{(j)}$ & exp data      & NNLO predictions & NLO predictions \\ \hline
    All data equal weight & 4618   & 0.1569±0.0041 & 0.1436±0.0042                & 0.1579±0.0052               \\
    DY         & 2100   & 0.1771±0.0060 & 0.1484±0.0047                & 0.1732±0.0070               \\
    NC DIS     & 2100   & 0.1570±0.0043 & 0.1439±0.0042                & 0.1626±0.0059               \\
    CC DIS     & 989    & 0.1729±0.0039 & 0.1542±0.0041                & 0.1646±0.0058               \\
    Direct $\gamma$ production     & 53     & 0.2322±0.0096 & 0.2386±0.0084                & 0.2447±0.0084               \\
    Top-quark pair production      & 66     & 0.2355±0.0072 & 0.2085±0.0063                & 0.2302±0.0085               \\
    Inclusive jet production       & 356    & 0.1899±0.0056 & 0.1626±0.0043                & 0.1785±0.0065               \\
    Di-jet production    & 144    & 0.2083±0.0073 & 0.1819±0.0055                & 0.203±0.012                 \\
    Single top production & 17     & 0.2344±0.0070 & 0.2252±0.0089                & 0.244±0.011                 \\
  \end{tabular}
  \caption{An overview of the $\varphi$ values defined in Eq.~(\ref{eq:phi}) for the different variations of fits performed for this study. The first column denotes the process for which the corresponding dataset has been given an increased weight in the fit. For each process we show the total number of datapoints in the corresponding datasets, and the $\varphi$ values for fits to the experimental data (left), fits to level-2 closure test data produced using NNLO calculations and fitted using NNLO theory (middle), and fits to level-2 closure test data produced using NNLO calculations and fitted using NLO theory (right).}
  \label{tab:phivalues}
\end{table}

As discussed before, increases in uncertainty do not necessarily have to be entirely, or even partially, due to inconsistencies between experimental datasets. To test this, we repeat the exercise by performing the fit to perfectly consistent closure test data. Any additional contribution to the uncertainty observed in weighted fits in this setting, cannot be attributed to inconsistencies between experimental measurements.
The results of the fits in this context are shown in column four of Tab.~\ref{tab:phivalues}. By comparing columns three and four it can be observed that the increase in PDF uncertainties between the unweighed fit and the weighted variations are qualitatively similar for the experimental setting and the closure test setting, though quantitatively the amounts by which the uncertainties increase is generally lower in the closure test setting. This means that while part of the increased uncertainties observed in the experimental setting are also observed in the closure test setting, another significant contribution is still unaccounted for.

Let us now consider another possible source of PDF uncertainty, namely, the theoretical uncertainty related to missing higher order corrections.
To test this we can again estimate the impact of missing higher order corrections using closure test data. Though this time in a slightly different setup, that is, one in which closure test data has been generated using NNLO theory calculations, but fitted using calculations at NLO.
The results of fits performed in this setting are shown in the fifth and last column of Tab.~\ref{tab:phivalues}. By comparing the results of the experimental setting and the fits to NNLO closure test data using NLO theory to generate predictions during the fit, it can be seen that the increase in $\varphi$ is at one-sigma agreement for all processes except inclusive jet production.

From this last comparison it can be concluded that any contribution to the PDF uncertainty due to inconsistencies between datasets in the NNPDF4.0 global dataset is at the level of the statistical fluctuations, it is thus negligible at the current level of PDF uncertainties.

\acknowledgments
The author would like to thank the members of the NNPDF collaboration for many insightful discussions during these studies. In particular Zahari Kassabov, Stefano Forte and Juan Rojo for comments at various stages of the project.
The author is supported by the European Research Council under the European Union’s Horizon 2020 research and innovation Programme (grant agreement number 740006).

\FloatBarrier
\bibliographystyle{JHEP}
\bibliography{blbl}

\end{document}